\documentclass[9pt,twocolumn,twoside]{opticajnl}
\journal{opticajournal} 

\setboolean{shortarticle}{true}

\usepackage{lineno}
\usepackage{siunitx}

\title{Agile laser wavelength tuning using dynamic targeting}

\author[1, *]{Robbe de Mey}
\author[2]{Spencer W. Jolly}
\author[3,4]{Alexandre Locquet}
\author[1]{Martin Virte}

\affil[1]{Brussels Photonics Team (B-PHOT), Vrije Universiteit Brussel, Pleinlaan 2, 1050 Brussel, Belgium}
\affil[2]{Service OPERA-Photonique, Université libre de Bruxelles, Brussels, Belgium}
\affil[3]{Georgia Tech – CNRS IRL 2958, Georgia Tech Europe, 2 Rue Marconi, 57070 Metz, France}
\affil[4]{School of Electrical and Computer Engineering, Georgia Institute of Technology, Atlanta, Georgia, 30332-0250, USA}

\affil[*]{robbe.de.mey@vub.be}

\begin{abstract}
Tunable lasers are essential and versatile tools in photonics, with applications spanning telecommunications, spectroscopy, and sensing. Advancements have aimed to expand tuning ranges, suppress mode hopping, and enable photonic integration. In this work, we explore the adaptation of dynamic targeting, a technique originally developed to stabilize lasers under optical feedback, as a method for achieving agile, fast, and continuous wavelength tuning. By adjusting the feedback rate and phase, we enable a stable and controlled frequency shift. We experimentally demonstrate reliable and reproducible tuning over 2.1 GHz using a free-space optical setup. Simulations further suggest that this approach could extend the tuning range to tens of GHz, with a potential scan speed exceeding $10^{17}$ Hz/s. These results highlight dynamic targeting as a promising route toward agile frequency control in semiconductor lasers.
\end{abstract}

\setboolean{displaycopyright}{false} 

\begin{document}
\maketitle

\noindent While lasers are ubiquitous, the ability to precisely tune their wavelength remains a subject of ongoing interest, driven by the ever-increasing demands of emerging applications. Moreover, integration of such a tunable laser on-chip now represents a crucial feature as it would pave the way towards a significant reduction of the size, cost, and efficiency of such lasers. Record tuning ranges exceeding 70 nm have been achieved with monolithically integrated lasers employing three cascaded asymmetric Mach-Zehnder interferometers~\cite{Latkowski2015}. The main drawback being the difficulty to gain fine and repeatable control of the laser wavelength \cite{Pajkovic2019, Skenderas2023}. On the other hand, ultrafast continuous tuning, up to \SI{12e15}{Hz/s}, has been demonstrated though over a much shorter wavelength range of about 500 MHz. The authors used hybrid integration, where an InP distributed feedback (DFB) diode laser was butt-coupled to a heterogeneous Si\textsubscript{3}N\textsubscript{4}–LiNbO\textsubscript{3} platform~\cite{Snigirev2023}. In the LiNbO\textsubscript{3} chip, a microresonator filters and feeds part of the light back into the laser. The wavelength tuning is then achieved through control of the microresonator frequency when operating the system in the so-called self-injection-locked or extended cavity regime.\\
While the extended cavity regime - corresponding to strong feedback - can be used to stabilize and control the laser, optical feedback is known to have a wide range of effects on semiconductor laser from linewidth reduction~\cite{Levine1995,Jin2021}, to instabilities, self-pulsing or even chaos. Feedback leads to the emergence of external cavity modes (ECMs), steady-state solutions whose frequencies are shifted relative to the solitary laser mode. As the feedback increases, the number of ECMs grows, and their coexistence becomes a fundamental factor influencing laser dynamics. Among these modes, the maximum gain mode (MGM) is particularly significant. By careful alignment of the external cavity it's possible to adjust the feedback phase so that there exists a stable mode at the MGM detuning~\cite{Levine1995}. To systematically retrieve and maintain the MGM, and thus remain stable at high feedback levels, a technique known as dynamic targeting was developed~~\cite{Wieland1997, Hohl1998, Heil2000, Ohtsubo2013}. It enables control over the laser's operating mode by adjusting both the feedback rate and phase, thus making it possible to keep the laser stable even at high feedback rates~\cite{Wieland1997}. This was confirmed experimentally by controlling the feedback rate and the laser current, thus adjusting the phase by tuning the lasing wavelength~\cite{Hohl1998}.\\
In this work, we explore how optical feedback could be further harnessed as a versatile solution to reach agile wavelength tuning without preventing on-chip integration of the system. We show that dynamic targeting can be leveraged for continuous wavelength tuning as increasing the feedback rate also shifts the MGM's wavelength. By dynamically adjusting the feedback phase as a function of the feedback rate, we can thus tune the laser wavelength through variation of the feedback rate while keeping its emission stable. We make an experimental proof-of-principle demonstration using a free-space setup with mechanical actuators, hence confirming the feasibility of this method for agile wavelength tuning. We obtain an excellent qualitative agreement with numerical simulations which we then use to evaluate the potential of this technique for agile wavelength tuning: we report a potential range of tens of GHz and a scan speed exceeding \SI{1e17}{Hz/s}.\\
\begin{figure}[ht]
	\centering
	\includegraphics[width = 0.8\linewidth]{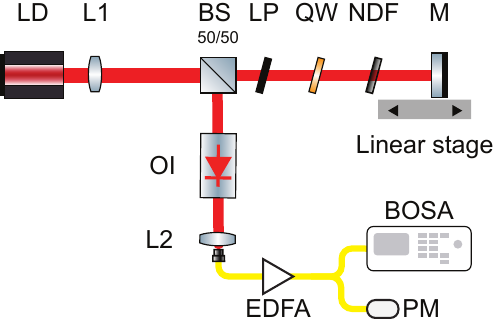}
	\caption{Experimental setup. LD: Laser Diode, L: Lens, BS: BeamSplitter, LP: Linear Polarizer, QW: Quarter Waveplate, NDF: Neutral Density Filter, M: Mirror, OI: Optical Isolator, EDFA: Erbium-Doped Fiber Amplifier, PM: Power Meter, BOSA: Brillouin Optical Spectrum Analyzer. Red indicates free space, and yellow indicates single-mode fibers.}
	\label{fig:figure1}
\end{figure}
\noindent Experimentally, we use the setup shown in Fig. \ref{fig:figure1}. The laser is a commercially available, single-mode, edge emitting DFB laser (3SP Technologies, 1953 LCV1), maintained at a constant temperature of $25^{\circ} C$. For an injection current of \SI{60}{mA}, just below three times the threshold current, the lasing wavelength is \SI{1551.5}{nm}. Mirror M is mounted on a high-precision linear stage (Newport XMS50) that can move over a 50 mm range with nanometer-scale precision. The mirror is positioned at \SI{110}{mm}. The feedback phase is adjusted by moving the position of the mirror at the sub-wavelength scale. Mechanical vibrations cause slight, unintended movements of the linear stage. However, by monitoring its position, we estimate that background mechanical vibrations lead to a mirror displacement of about $\pm \SI{25}{nm}$, which introduces a maximum phase error of \SI{25}{\nm}/(\SI{1551.5}{\nm}/2) = 3.2\% at \SI{1551.5}{nm}. The feedback rate is controlled in two stages: a neutral density filter (NDF) wheel is used to set the initial feedback strength, the feedback strength is dynamically adjusted using a linear polarizer (LP) and a quarter-wave plate (QWP) mounted on a motorized rotation stage (Thorlabs PRM1/MZ8E). The light passes through the QWP twice, inducing a rotation of the linearly polarized laser light emitted by the laser. The feedback light is then partially attenuated by the LP in function of the rotation angle. With this technique, there is a $\cos^2$ dependence between the feedback strength and the rotation angle of the QWP. The tuning method works by adjusting both the feedback rate and phase. Therefore, we adjust the angle such that the feedback rate is controlled linearly.\\
Numerically, to model our semiconductor laser with optical feedback, we use the normalized Lang and Kobayashi equations which read as follows \cite{Lang1980,Rogister1999}:
\begin{align}
	\label{eq:H2LKOneDelay}
	\dot{E}(t) &= \frac{1}{2}(1-i\alpha)N(t)E(t)+ \kappa E(t-\tau)\exp(i \phi),\\
	\dot{N}(t) &= (P-N(t)-(1+2N(t))E^2(t))/T,
\end{align}
\noindent where \textit{E} is the complex electric field and \textit{N} represents the carrier density. The laser parameters are the linewidth enhancement factor $\alpha = 3$, the carrier lifetime $T = 1000$, and the pump current $P = 1$ (corresponding to twice the lasing threshold). The equations are normalized with respect to the photon lifetime. For these values, the relaxation oscillation period is 141 \cite{demey2024}. The feedback parameters are the feedback rate $\kappa$, the external cavity delay $\tau$, and the feedback phase $\phi$. From a physical point of view, the phase is equal to $\omega_0\tau$, where $\omega_0$ is the angular frequency of the solitary laser. Here, we, however, consider that sub-wavelength variations of the feedback length do not impact the delay itself but only lead to a change in the feedback phase, and thus control $\phi$ independently from $\tau$.\\
Without feedback, the laser operates at a single fixed point, corresponding to the natural mode of a solitary laser. When optical feedback is introduced, ECMs emerge, shifting the lasing wavelength away from that of the solitary laser. The maximum gain mode (MGM) is, interestingly, the one showing the largest shifting in terms of frequency and cannot be destabilized by increasing feedback \cite{Levine1995}. In Ref. \cite{Levine1995}, the authors provide an analytical description of the MGM for $\phi = \tau \alpha \kappa$. This equation shows that keeping the laser on the MGM requires an interplay between the feedback phase $\phi$ and feedback rate $\kappa$. The ratio between the two values being imposed by the delay $\tau$ and the linewidth enhancement factor $\alpha$. The lasing wavelength shift follows the equation $\Delta \omega = -\kappa \alpha$, linking the tuning range to the $\alpha$ factor and the feedback rate. The idea behind the tuning mechanism we propose is therefore based on an increase of the feedback rate---inducing a wavelength shift---combined with a controlled variation of the feedback phase to keep the laser on the MGM. It is interesting to note that the tuning range does not depend on the time-delay, but only on the maximal feedback rate that can be achieved and the $\alpha$-factor.\\
\begin{figure*} [h]
	\centering
	\includegraphics[width = \linewidth]{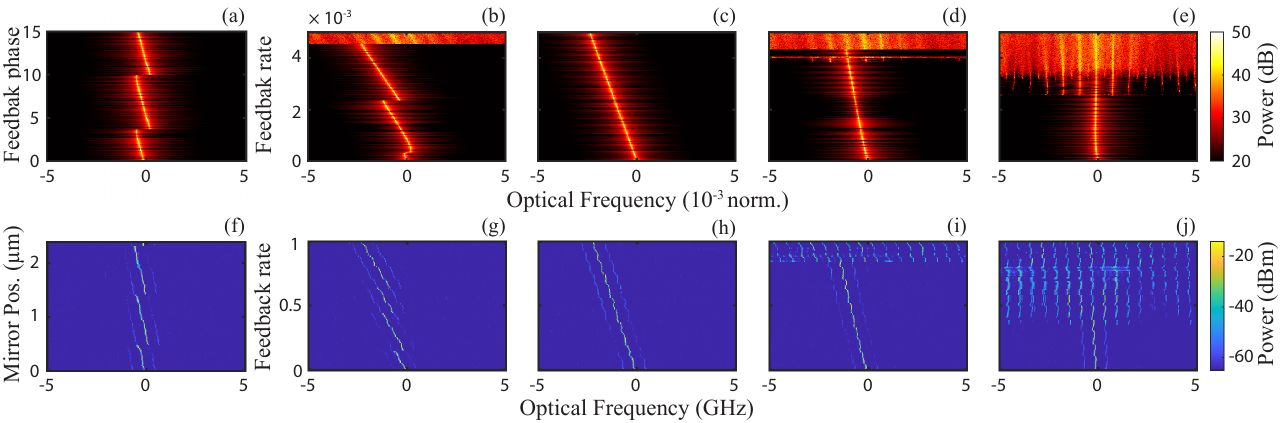}
    \caption{Optical spectrum variations as a function of feedback parameters. \textbf{(a)-(e) Simulation results:} (a) Sweeping only the feedback phase. (b) Feedback phase changing twice as fast as the feedback rate. (c) Tuning of feedback rate and phase using the ideal ratio. (d) Feedback phase changing at half the speed of the feedback rate. (e) Sweeping only the feedback rate ($\kappa$). The parameters used are $\alpha = 3$, $P = 1$, and $\tau = 1000$. \textbf{(f)-(j) Experimental results:} (f) Sweeping only the feedback phase. (g)-(i) Tuning with different tuning ratios: (g) phase tuned faster (h) ideal case, and (i) phase tuned slower (j) Sweeping only the feedback rate.}
	\label{fig:figure2}
\end{figure*}
To validate the tuning method, we conduct a series of simulations, exploring variations of different parameters to assess their impact on wavelength control. We first examine the effect of sweeping only the feedback phase while keeping the feedback rate fixed at an intermediate value of $\kappa = 0.001$ , shown in Fig. \ref{fig:figure2} (a). The wavelength undergoes tuning, but jumps back periodically, indicating that the laser does not remain continuously locked to the MGM. In Fig. \ref{fig:figure2} (b), (c), and (d), both feedback rate and phase are tuned synchronously with a different ratio between them. Deviations from the ideal proportionality coefficient lead to instabilities, with the laser either undergoing mode hopping, as seen in Fig. \ref{fig:figure2} (b), or exhibiting chaotic fluctuations, see Fig. \ref{fig:figure2} (d). In Fig. \ref{fig:figure2} (c), we tune the feedback parameters with the ratio given by $\phi = \tau\alpha\kappa $. Unlike the previous cases, the laser output remains stable, and the wavelength tuning proceeds smoothly without mode hops. The lasing wavelength matches that of the MGM predicted when simulating the fixed points, confirming that the system remains in the MGM throughout the tuning process~\cite{DemeyPhD2025}. Finally, as a last comparison point, we sweep the feedback rate $\kappa$ from 0 to 0.005 without adjusting the feedback phase in Fig. \ref{fig:figure2} (e). We observe that the system transitions from a stable state to oscillatory behavior and eventually enters a chaotic regime, confirming the destabilizing effect of increasing feedback rate without phase compensation.\\
While simultaneous and continuous tuning is used numerically, an iterative step-by-step adjustment of both the feedback rate and phase sequentially is far easier to achieve experimentally. To avoid inducing instabilities in the tuning, we use small steps of maximum \SI{50}{nm} when changing the mirror position. In addition, determining the ideal ratio between feedback rate and phase is significantly more challenging than in the simulations. We therefore proceeded through trial and error. The experimental results are shown in the second row of Fig.~\ref{fig:figure2}. In Fig.~\ref{fig:figure2} (f), only the feedback phase is varied. Like in the simulation, the lasing wavelength is swept but exhibits periodic jumps. The results of sweeping both the feedback rate and phase are shown in Figs.~\ref{fig:figure2} (g), (h), and (i). The full accessible range of feedback rates is covered for each case, but the total displacement of the mirror is adjusted to explore different tuning ratios between the feedback rate and phase. The optimal ratio, where the lasing wavelength is tuned over a range of \SI{2.1}{GHz} and for which the laser remains on the same mode along the whole range, is shown in Fig.~\ref{fig:figure2} (h). The tuning range is limited by the accessible range of feedback strength, as the QWP is rotated from minimal to maximal transmission of the light in the external cavity. In Fig. \ref{fig:figure2} (g), the feedback phase is tuned faster than the feedback rate, leading to hops in the spectrum. On the other hand, in Fig. \ref{fig:figure2} (i), the feedback phase is swept slower which leads to instabilities for strong feedback. Finally, as a comparison point, we show in Fig.~\ref{fig:figure2} (j) the case where only the feedback rate is swept. As expected, the laser initially operates on a stable state but soon becomes unstable, resulting in a broader spectrum. We observe excellent qualitative agreement for all these cases between the experimental and numerical results.\\
While hysteresis has been reported when varying the feedback phase alone~\cite{Ohtsubo2013}, we observed no hysteresis when the laser remains on the same mode. As a result, tuning by dynamic targeting can be used back and forth without significant issues (Fig.~\ref{fig:figure3}). At the end, the laser emission return to its initial wavelength, demonstrating bidirectional tunability. On the other hand, we also observe oscillations or ‘wiggling’ of the lasing wavelength during tuning, reducing the linearity of the tuning. The origin of this remains unclear.\\
\begin{figure}[h]
	\centering	
	\includegraphics[width = \linewidth]{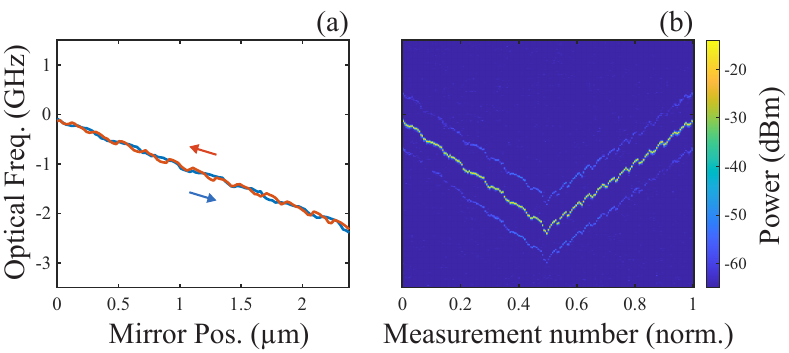}
    \caption{Experimental bidirectional tuning of the lasing wavelength. (a) Lasing wavelength as a function of mirror position (and feedback rate). Blue: forward tuning, orange: backward tuning. (b) Optical spectrum map illustrating the wavelength shift.}
	\label{fig:figure3}
\end{figure}
\begin{figure*} [h]
	\centering
	\includegraphics[width = \linewidth]{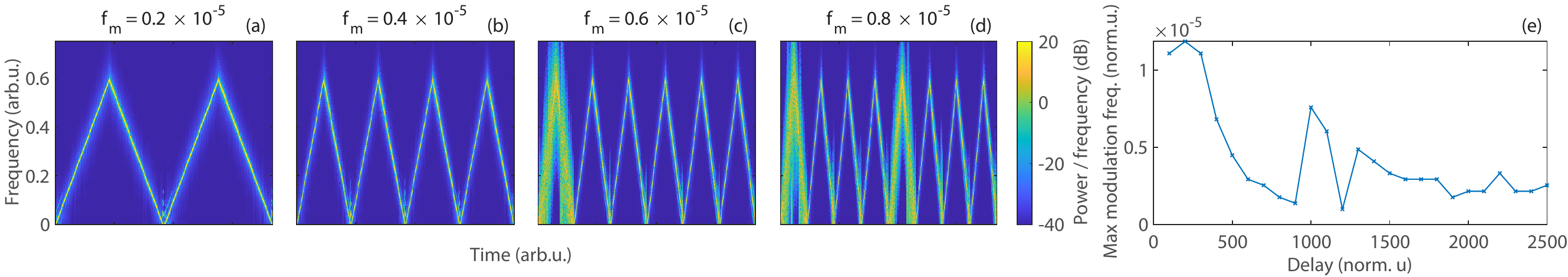}
    \caption{ (a)-(d) Simulation, of the tuning the lasing wavelength at high speed with $\tau = 1000$. The optical spectrum is plotted versus the time. From (a) to (d) the frequency of the modulation frequency ($f_m$) is increased. (c) One roundtrip is needed before the transient settles down. (d) At a certain modulation frequency, the system tuning mechanism breaks down. (e) Simulation of the maximum modulation frequency when changing the delay.}
	\label{fig:figure4}
\end{figure*}

\noindent We now use simulations to investigate the limits of this tuning method, and especially the sweeping speed. A triangular signal of frequency $f_m$ is applied to the feedback rate and phase. The feedback rate $\kappa$ is modulated between 0 and 0.2 while the feedback phase $\phi$ is set as $\phi = \tau \alpha \kappa$. We then increase the modulation frequency and examine how fast the system can respond to this signal. In Fig. \ref{fig:figure4} (a-d), we show evolution of the laser optical spectrum versus time for $\tau=1000$ and $f_m = 0.2$, $0.4$, $0.6$ and $0.8\times 10^{-5}$ respectively. Like the rate equations used, the frequency is normalized by the photon lifetime. For slower modulation frequency, the system follows well the triangular signal and the laser remains on the MGM. For $f_m = \SI{0.6e-5}{}$, we however observe a first issue: the wavelength follows the same pattern but with a much broader spectrum. Then, at the second period, the laser seems to have successfully reached the MGM. At $f_m = \SI{0.8e-5}{}$, the laser exhibit the same behavior but the same broader spectrum appears on the 5th period indicating that the laser can escape from the MGM. We extend these simulations to different delays to estimate the maximum tuning speed. In each case, we retain the largest modulation frequency $f_m$ for which the laser remains on the MGM for several periods. The results are presented in Fig. \ref{fig:figure4} (e). For a delay of 1000, a value corresponding to the long-cavity regime, the system successfully tracks the modulation up to a frequency of \SI{8.37e-6}. Interestingly, the relationship between the delay and the maximum modulation frequency is non-monotonic. The case with a delay of 1000 exhibits better tracking of higher modulation frequencies than the case with a delay of 500, highlighting the intricate interplay between feedback dynamics and modulation response. These results indicate that while the delay does not directly affect the achievable tuning range, it plays an important role in determining the dynamical response and the system’s ability to track rapid modulations. Nevertheless, the short cavity is likely the most promising configuration to achieve fast tuning.\\
For the sake of comparison, we can estimate the sweeping speed in SI units by de-normalizing the different parameters. All equations are normalized in time by the photon lifetime $\tau_p$ which is typically around 5 ps for a common semiconductor laser. We successfully tune the laser wavelength up to $\Delta f = -\kappa \alpha = -0.6$ with $\alpha = 3$ and a maximal value of $\kappa=0.2$, which corresponds to $\Delta f/\tau_p = - 120 \, GHz$. In terms of modulation speed, $f_m > 0.2 \times 10^{-5}$ can be reached for most configurations while $10^{-5}$ appears to be within reach with shorter delays. In this case, the laser can be tuned over the full tuning range of $120\, GHz$ in half a period $\tau_P/(2 f_m)= 1.25 \, \mu s$ and $0.25 \, \mu s$ respectively. This yields a scan speed of \SI{96e15}{Hz/s} and \SI{480e15}{Hz/s}, more than one order of magnitude above the \SI{12e15}{Hz/s} speed reported in~\cite{Snigirev2023}.

In conclusion, the results of this study demonstrate the viability of dynamic targeting as a method for precise, fast, and continuous wavelength tuning. Originally developed to stabilize lasers under strong feedback, this technique enables tuning by adjusting both the feedback rate and phase, ensuring stable operation while shifting the lasing wavelength. Experimentally, we validated this tuning approach and demonstrated that the method enables precise and reproducible tuning over a range of \SI{2.1}{GHz}. Through rate equation simulations, we have shown that this method can achieve high-speed tuning, with simulated scan speeds potentially exceeding $10^{17}$ Hz/s. The ability to control the lasing wavelength with minimal perturbation makes this approach a strong candidate for integration into photonic integrated circuits (PICs), especially considering the potential performance improvement linked with the use of shorter external cavities. Nevertheless, some limitations remain to be considered. The laser must either start close to the desired mode as improper initialization can hinder stable tuning. The initial feedback phase must also be carefully set to ensure operation on the correct mode. Next, the tuning process obviously requires precise synchronization of feedback parameters, especially if high scan speed are targeted, as any deviation can lead to instability or mode hopping. Furthermore, the dependence on optical feedback introduces potential vulnerabilities to parasitic reflections, which could disrupt the system and degrade performance.\\
Future work will focus on on-chip implementation. The free-space optical setup served as an interesting proof-of-principle demonstration, but its reliance on mechanical actuators limits both speed, precision, and tuning range compared to what could be achieved with an integrated system. With continued development, this method has the potential to become a key enabling technology for next-generation high-speed tunable laser systems.

\begin{backmatter}
\bmsection{Funding}
The authors acknowledge the financial support of Fonds Wetenschappelĳk Onderzoek (FWO), embassy of France in Belgium, TOURNESOL program, the Région Grand Est, the French PIA project “Lorraine Université d’Excellence” (reference ANR-15-IDEX-04-LUE), and the Fonds De La Recherche Scientifique (FNRS).


\bmsection{Disclosures} The authors declare no conflicts of interest.

\bmsection{Data Availability Statement}The data produced in this study have been deposited in the open access digital repository Zenodo and can be accessed at \href{https://doi.org/10.5281/zenodo.15583123}{https://doi.org/10.5281/zenodo.15583123}. 

\end{backmatter}

\bibliography{biblo.bib}

\end{document}